\documentclass[aps,prc,twocolumn,groupedaddress,showpacs]{revtex4}
\usepackage{graphicx}

\begin{document}
\title{Airy structure in $^{16}$O+$^{14}$C nuclear rainbow scattering 
 }

\author{
S. Ohkubo$^{1}$ and  Y. Hirabayashi$^2$  
 }
\affiliation{$^1$ Research Center for Nuclear Physics, Osaka University, 
Ibaraki, Osaka 567-0047, Japan }
\affiliation{$^2$Information Initiative Center,
Hokkaido University, Sapporo 060-0811, Japan}

\date{\today}

\begin{abstract}
 The Airy structure in $^{16}$O+$^{14}$C rainbow scattering is studied 
with an extended double folding 
(EDF) model that describes all the diagonal and off-diagonal coupling potentials 
derived from  the microscopic realistic  wave functions for $^{16}$O  
  using  a density-dependent   nucleon-nucleon force.
The experimental angular distributions at $E_L$=132, 281 and 382.2 MeV are well reproduced
 by the calculations. By studying the energy evolution of the Airy structure, the 
 Airy minimum at around $\theta$=76$^\circ$ in the angular distribution at $E_L$=132 MeV 
is assigned as the second order Airy minimum $A2$ in contrast to the recent literature
 which assigns it as the third order $A3$. 
 The Airy minima  in the 90$^\circ$ excitation function is investigated in comparison 
with   well-known  $^{16}$O+$^{16}$O and $^{12}$C+$^{12}$C systems. 
 Evolution of     the  Airy structure 
 into the molecular resonances with the 
$^{16}$O+$^{14}$C cluster structure  in the low  energy region around $E_{c.m.}$=30 MeV
 is  discussed. 
 It is   predicted theoretically for the first time  for a non-$4N$ $^{16}$O+$^{14}$C system   
that   Airy elephants    in the 90$^\circ$ excitation function are present.
\end{abstract}

\maketitle

\par 
\section{INTRODUCTION}
  Nuclear rainbow scattering,
which is observed under incomplete absorption, can uniquely determine the interaction 
potential family  up to the internal region  without ambiguity \cite{Khoa2007}.
The interaction potential for the  $^{16}$O+$^{16}$O system  has been most thoroughly investigated both
 experimentally and theoretically.
Although a shallow potential had been used in heavy ion scattering and reactions 
for many years \cite{Hodgson1978}, the observation
of a nuclear rainbow in  $^{16}$O+$^{16}$O scattering at $E_L$=350 MeV finally 
showed that a global  interaction potential for this system is deep \cite{Khoa2007}.
 It was shown in Ref.\cite{Ohkubo2002}  that the global  deep potential
determined in nuclear rainbow scattering can describe in a unified way not only the  
  prerainbows and   the  Airy structure  in the
 90$^\circ$ excitation function, but also the molecular resonances in the low energy region 
and  the band structure  with  the $^{16}$O+$^{16}$O  cluster structure.
 It was also found \cite{Ohkubo2002} that the   highest order Airy
 structure \cite{Michel2001} evolves   into molecular  resonances with  the 
  $^{16}$O+$^{16}$O cluster structure in $^{32}$S as the incident  energy decreases.
 The  gross structures   in the  90$^\circ$ excitation function in rainbow scattering  
separated by the Airy minima have been visually interpreted as panchdermous 
 Airy elephants in Ref. \cite{McVoy1992}.

\par
Rainbow scattering and interaction potentials for the asymmetric  $^{16}$O+$^{12}$C system
 have been 
studied systematically at $E_L$=63-260 MeV \cite{Nicoli2000,Szilner2001,Ogloblin1998,Ogloblin2000} and 
 $E_L$=608-1503 MeV \cite{Khoa1994} and  
a global potential was determined. The global potential could  explain  not only the
 rainbows and  prerainbows 
\cite{Nicoli2000,Szilner2001,Ogloblin1998,Ogloblin2000,Khoa1994,Khoa2007}
but also the  molecular resonances in the low energy region and 
the superdeformation with the  $^{16}$O+$^{12}$C cluster structure  
    in a unified way \cite{Ohkubo2004B}.
However, in order to explain 
the    Airy minimum observed at  much larger angles at around $E_L$=300 MeV 
 \cite{Ogloblin2003}, which was   impossible to reproduce 
in the optical model calculations  with the   global potential,
a deeper family potential was needed.
  In Ref.\cite{Ogloblin2003} the order of the Airy minimum was reassigned 
systematically to be one  higher than that in   previous literature in Refs.
\cite{Nicoli2000,Szilner2001,Ogloblin1998,Ogloblin2000,Khoa1994,Khoa2007}. For example,
 the Airy minimum at $\theta$=82$^\circ$ at  $E_L$=132 MeV was assigned $A3$
 instead of  $A2$.
Very recently this  dilemma  was rescued \cite{Ohkubo2014} by 
noticing that the  Airy minimum at the large angle
is   a new kind of  Airy minimum 
 caused dynamically by the coupling  to an  excited state of $^{12}$C 
and it was found that the experimental angular distributions are   
 reproduced by the coupled channel calculations 
 with a global extended folding  potential derived from   the  microscopic wave 
 functions for  $^{12}$C and $^{16}$O.

\par 
The  $^{16}$O+$^{14}$C system is situated between   $^{16}$O+$^{16}$O and 
 $^{16}$O+$^{12}$C. Ogloblin {\it et al.} \cite{Ogloblin2003}
 measured rainbow scattering for the 
 $^{16}$O+$^{14}$C system at 132, 281 and 382.2 MeV.  Glukhov {\it et al.} \cite{Glukhov2007}
investigated the Airy structure and  concluded   that the order of the
Airy minimum at  $\theta$=76$^\circ$ in the angular distribution at  $E_L$=132 MeV 
  is $A3$, which is similar to the Airy minimum $A3$ at   $\theta$=82$^\circ$ in the angular
 distribution of $^{16}$O+$^{12}$C   at $E_L$=132 MeV  claimed with a deeper family 
potential in  Ref.\cite{Ogloblin2003}.

\par
The purpose of this paper is to study the Airy structure of rainbow scattering for
 the $^{16}$O+$^{14}$C  system with the extended double folding 
 model used successfully in Ref.\cite{Ohkubo2014} for the  $^{16}$O+$^{12}$C system
 and to  determine  the  order of the Airy minimum  from the energy evolution of the 
 Airy minimum over a wide range of incident energies.
 It is shown that the Airy minimum at  $\theta$=76$^\circ$ in the angular
 distribution at $E_L$=132 MeV  is  $A2$. This is  
   different from the previous assignment in Ref.\cite{Glukhov2007}.
   The evolution of the Airy structure   into the molecular
 resonances and the cluster  structure  in  the low energy region is discussed 
 in comparison with typical system such as  $^{16}$O+$^{16}$O.

\section{EXTENDED DOUBLE FOLDING MODEL}
\par
We study   rainbow scattering for  $^{16}$O+$^{14}$C  with an extended double folding 
(EDF) model that describes all the diagonal and off-diagonal coupling potentials 
derived from  the microscopic   wave functions for $^{16}$O  
  using  a density-dependent   nucleon-nucleon force.
  The diagonal and coupling potentials 
for the $^{16}$O+$^{14}$C system are calculated using the EDF  model
 and are given as follows:
\begin{eqnarray}
\lefteqn{V_{ij}({\bf R}) =
\int \rho_{ij}^{\rm (^{16}O)} ({\bf r}_{1})\;
     \rho_{00}^{\rm (^{14}C)} ({\bf r}_{2})} \nonumber\\
&& \times v_{\it NN} (E,\rho,{\bf r}_{1} + {\bf R} - {\bf r}_{2})\;
{\it d}{\bf r}_{1} {\it d}{\bf r}_{2} ,
\end{eqnarray}
\noindent where $\rho_{00}^{\rm (^{14}C)} ({\bf r})$ represents the diagonal 
 nucleon density of the ground state of $^{14}$C, which is obtained by the convolution of 
the proton size from the charge density distribution taken from  Ref.\cite{DeVries1987}.
$\rho_{ij}^{\rm (^{16}O)} ({\bf r})$ is the diagonal ($i =j$) or transition
  ($i \neq j$) nucleon  density of  $^{16}$O 
  taken from  the microscopic $\alpha$+$^{12}$C  cluster model  wave functions calculated
 in  the orthogonality 
condition model (OCM) in Ref.\cite{Okabe1995}, which uses  a  realistic size parameter both 
 for the $\alpha$ particle and $^{12}$C. This is an extended version of
Ref.\cite{Suzuki1976}, which well reproduces almost  all the energy levels  
 up  to $E_x$$\approx$13 MeV and the  electric transition probabilities  in $^{16}$O. 
The wave functions have been  successfully used for the systematic 
analysis of elastic and inelastic  scattering   over a wide range of 
incident energies \cite{Hirabayashi2013,Ohkubo2014,Ohkubo2014B,Ohkubo2014C}.
 We take into account  the important transition densities 
available in Ref.\cite{Okabe1995}, i.e., g.s $\leftrightarrow$  $3^-$ (6.13 MeV) and 
$2^+$ (6.92 MeV)  in addition to all the  diagonal potentials.  
  For the  effective interaction   $v_{\rm NN}$     we use  
 the DDM3Y-FR interaction \cite{Kobos1982}, which takes into account the
finite-range nucleon  exchange effect.
  In the calculations we introduce the normalization factor  $N_R$ for 
 the real part of the  double folding potential \cite{Satchler1979,Brandan1997}. 
An imaginary potential with a  Woods-Saxon volume-type form factor (nondeformed) is
 introduced   phenomenologically to take into account the effect
of absorption due to other channels.

\section{AIRY STRUCTURE IN ELASTIC  $^{16}$O+$^{14}$C SCATTERING }
 \par In Fig.~1 the  angular distributions in elastic  $^{16}$O+$^{14}$C 
 scattering  calculated  using the single channel double folding (DF) model potential 
 are compared with the
experimental data  \cite{Ogloblin2003} at  $E_L$= 132, 281    and 382.2  MeV.
The normalization factor and  volume integral per nucleon pair, $J_V$, for the
real folding potential,  and the imaginary potential parameters used are given in Table I.
The experimental angular distributions are 
well reproduced by the single channel calculations. The calculated cross sections are 
decomposed into the farside (dashed line) 
and nearside (dash-dotted line) components. The nearside component decreases rapidly beyond the 
 diffraction region and the farside component dominates toward   the intermediate angular
 region.  Thus the  broad 
structure  of the angular distribution is the Airy structure of the  nuclear rainbow 
caused by refractive scattering. 

\begin{figure}  [t]
\includegraphics[keepaspectratio,width=8.7cm] {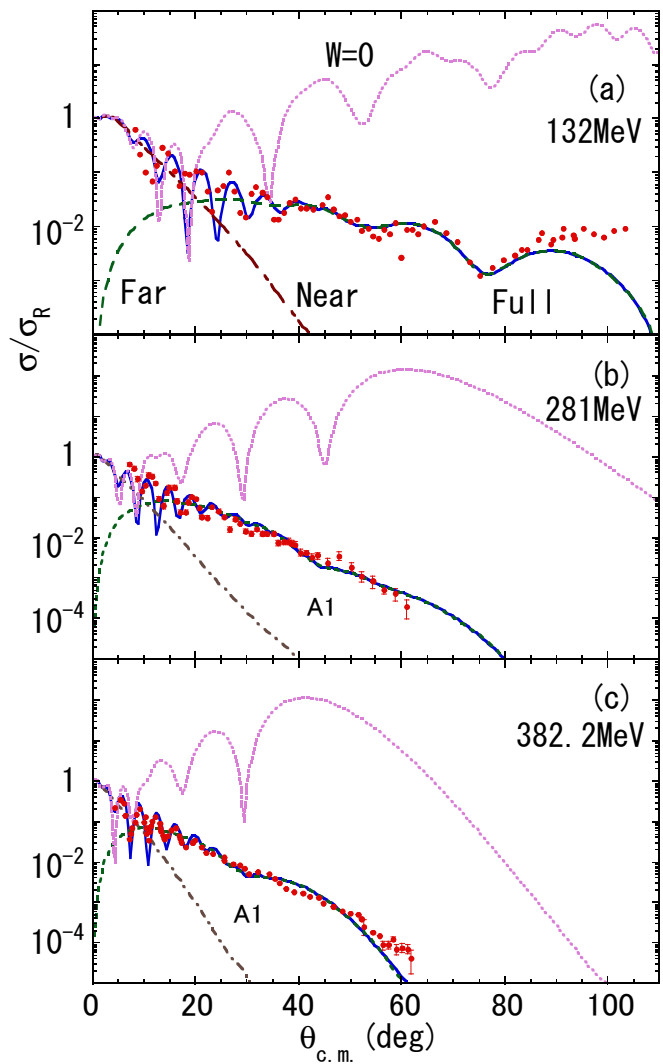}% Here is how to import EPS art
 \protect\caption{\label{fig.1} {(Color online) 
 Comparison of the  single channel  DF potential model calculations (blue solid line)
 with the  experimental 
angular distributions of cross sections (ratio to Rutherford  scattering) (points)
 \cite{Ogloblin2003}  in elastic  $^{16}$O+$^{14}$C  scattering
at $E_L$= 132 MeV (a), 281 MeV (b),  and 382.2  MeV (c).
The  dashed (green) and dash-dotted (grey) lines display the calculated farside and nearside
 components, respectively. The angular distributions calculated by switching off the imaginary
potential (W=0) are displayed by the dotted (pink) lines.  
}
}
\end{figure}

%Table I
\begin{table}[t]
\begin{center}
\caption{ \label{Table I}
The   volume integral per nucleon pair $J_V$  of the 
 the ground state diagonal potential (in units of MeVfm$^3$)  and the   imaginary potential parameters used in the
 single channel double folding calculations in Fig.~1 and coupled channel calculations   with EDF 
in Fig.~2. $N_R$=1 is used except $N_R=$0.95 at 132 MeV (single channel) and 281 MeV (coupled
 channel).
}
\begin{tabular}{cccccccccc}
 \hline
  \hline
$E_{L}$ & $J_V$      & $W$  &$R $ &$a $& &$J_V$  & $W $ & $R $ &$a $  \\
 \cline{2-5}    \cline{7-10}
 &      \multicolumn{4}{c}{(single channel cal.)}   &  &
\multicolumn{4}{c}{(coupled channel cal.)}           \\           
 \cline{2-5}    \cline{7-10}
(MeV) &  &(MeV) &(fm) &(fm) & & &(MeV)  & (fm) &(fm)   \\   
 \hline
 \hline
 132 & 285   &  17.0   & 5.60 & 0.70& &300 & 16.0 & 5.55  &0.50   \\
 281&  273     &  26.0 & 5.65 & 0.60 & &259& 22.0 & 5.60 &0.55    \\
 382.2 & 254     &  26.5  & 5.65 & 0.70 & &254& 26.0   & 5.45 &0.75\\
 \hline                          
 \hline                          
\end{tabular}
\end{center}
\label{Table1}
\end{table}

\par
The order of the Airy minimum is determined by  calculating  the angular 
 distribution  by switching off the imaginary 
potential at the highest energy $E_L$=382.2 MeV in Fig.~1(c). 
 The fall-off of the cross sections in the angular distribution, i.e., the darkside
 of the rainbow, starts beyond $\theta$=$40^\circ$, which means that the minimum at $30^\circ$ is the
 first order Airy minimum $A1$.  At $E_L$=281 MeV  in Fig.~1(b)  the 
second order Airy minimum $A2$ is seen at  $30^\circ$ in addition to the $A1$ at $45^\circ$.

%Fig.2
\par  In order to determine  the order of the Airy minimum at $\theta$=$76^\circ$ 
  at the lowest  energy $E_L=$132 MeV in Fig.~1(a) without ambiguity,   
 the energy evolution of the Airy  structure of the angular distribution between 
$E_L$= 281 MeV and 116 MeV is  calculated using the single channel double folding
 potential by switching off the imaginary potential. This is displayed in Fig.~2.  The value of $N_R$ was interpolated
or extrapolated from    those at $E_L$=281 MeV ($N_R$=1.0) and 132 MeV ($N_R$=0.95).
The energy dependence of the DF potential comes mostly  from  the DDM3Y-FR effective two
body interaction.  At $E_L$=140 MeV the $A1$ Airy minimum is clearly seen  at $100^\circ$.
 Thus the Airy minimum at $76^\circ$ at $E_L$=132 MeV is found to be  $A2$.
 This assignment of the $A2$  Airy minimum  at $76^\circ$ for 
 the $^{16}$O+$^{14}$C system at 132 MeV 
 corresponds  well to the  $A2$ assignment of the Airy minimum   at   $82^\circ$ 
  for the  $^{16}$O+$^{12}$C system  at the same $E_L$=132 MeV
 in Refs.\cite{Ogloblin1998,Ogloblin2000,Szilner2001}. 
The energy evolution of the Airy minimum seems to support the interpretation 
 that the minimum (not visible in Fig.~1) at around  $120^\circ$
 at $E_L$=132 MeV is a remnant of the  Airy minimum $A1$.  In fact, the  calculated angular distribution 
beyond this  angle turns into  diffraction-like high-frequency oscillations rising toward the
 extreme backward angle $180^\circ$.  
At $E_L=$132 MeV in Fig.~1(a) $A3$ is observed at $50^\circ$.

\par
It is worth mentioning that  coupled reaction channel (CRC) calculations for the $^{16}$O+$^{14}$C system at
 $E_L$=132 and 281 MeV in Ref.\cite{Rudchik2011} show that
 the potential scattering dominates at  angles  smaller than  $90^\circ$ and
  the contribution of the two proton cluster transfer reactions dominates at large angles.
The minimum at around    $\theta$=$120^\circ$ in $^{16}$O+$^{14}$C scattering
 at $E_L$=132 MeV visible in the farside component in Fig.~2  could  be seen only
as a remnant of the  Airy  minimum $A1$ in experiment.
We note  that the wrong  $A3$ assignment to  the Airy minimum 
at $76^\circ$ at 132 MeV in Ref.\cite{Glukhov2007}, which should be  $A2$ due to the Luneberg lens \cite{Michel2002}
 of the mean field potential,
  was done simply based on the similarity of  the shape of the  angular distributions 
and the   Airy minimum
 between $^{16}$O+$^{14}$C  and  $^{16}$O+$^{12}$C scatterings  at the same energy.

%Fig.2 energy evolution
\begin{figure}[t]
\includegraphics[keepaspectratio,width=8.6cm] {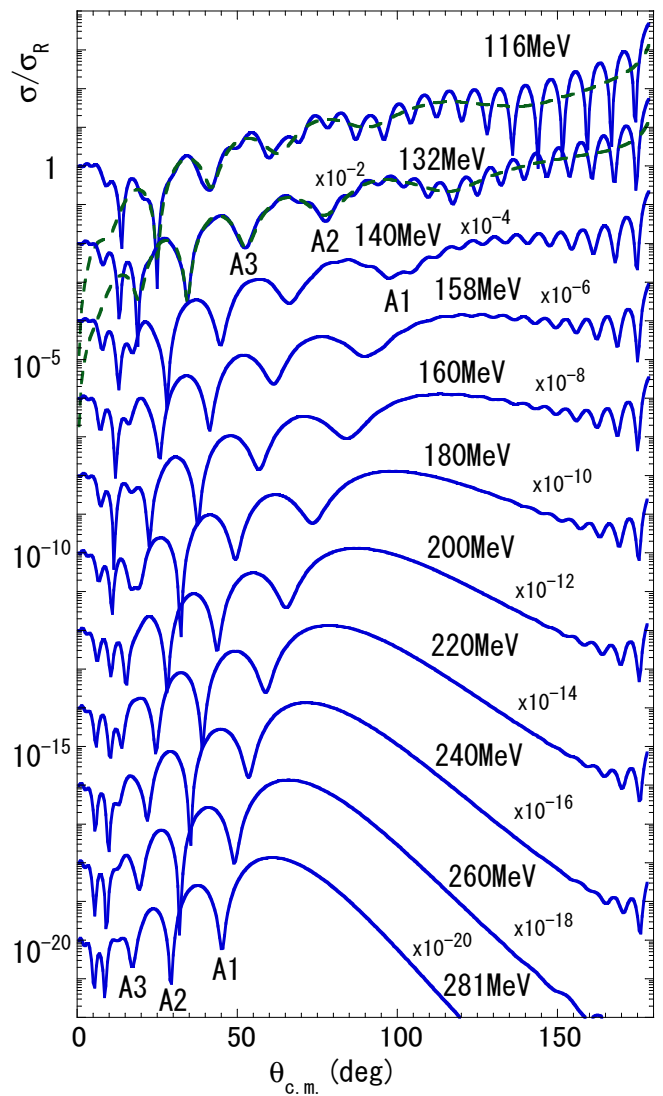}% Here is how to import EPS art
 \protect\caption{\label{fig.2} {(Color online) 
 The energy evolution of the Airy structure in the angular distributions of cross sections 
(ratio to Rutherford  scattering)  in $^{16}$O+$^{14}$C scattering calculated 
using the  single channel DF potential by switching off the imaginary potential is shown
 by the  solid lines. The dashed lines at 116 and 132 MeV are the  farside component
of the calculated cross sections.
}
}
\end{figure}

\begin{figure}[t]
\includegraphics[keepaspectratio,width=8.7cm] {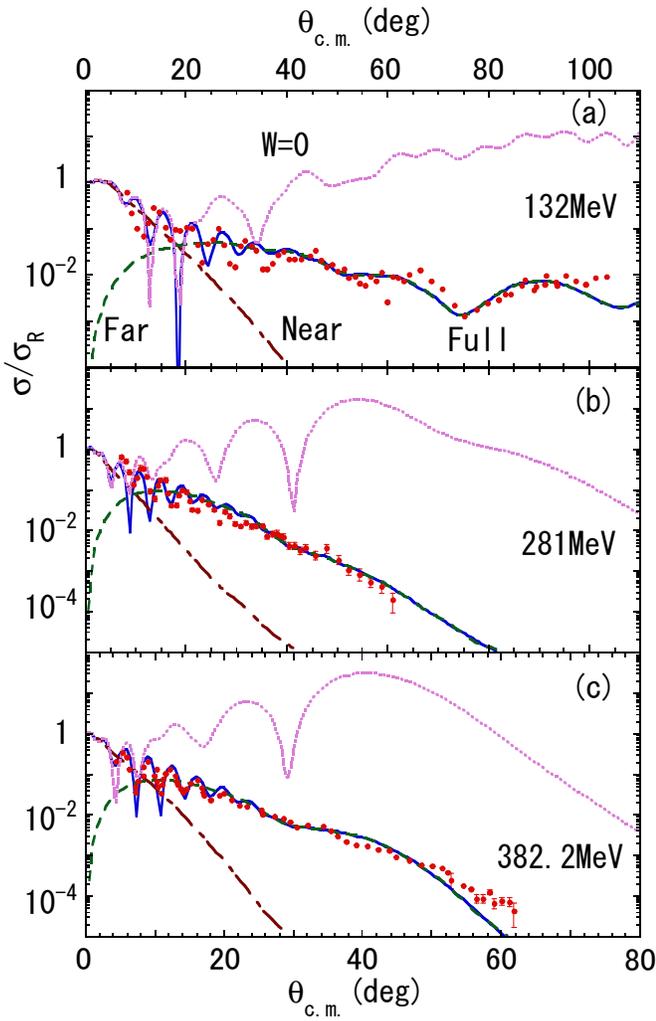}% Here is how to import EPS art
 \protect\caption{\label{fig.3} {(Color online) 
  Comparison of the coupled channel calculations (blue solid line)  
 with experimental 
angular distributions of cross sections (ratio to Rutherford  scattering) 
(points) \cite{Ogloblin2003}  in $^{16}$O+$^{14}$C  scattering
at $E_L$= 132 MeV (a), 281 MeV  (b),  and 382.2 MeV (c).
The  dashed (green) and dash-dotted (grey) lines display the calculated farside and nearside
 components, respectively. The angular distributions calculated by switching off the imaginary
potential (W=0) are displayed with   dotted (pink) lines. 
Note that the upper horizon scale is for (a) and (b) and the lower horizontal scale is for (c). 
}
}
\end{figure}

%Fig.3
\par
In Fig.~3 the angular distributions calculated using the coupled channel method
 are compared with the experimental data. The  potential parameters used are given in Table I.
The experimental angular distributions are  well reproduced by the coupled channel 
 calculations. There is little difference between the coupled channel and the single channel
 calculations in Fig.~1
   at the higher energies, 382.2 and 281 MeV, although     the Airy minimum $A1$ 
is slightly  shifted  forward at 132 MeV  compared with the single channel calculation. 
Essentially,  the effect of the channel coupling on the Airy structure is not important
 and the angular 
distributions are well described by the mean field  DF potential.  A dynamical secondary 
rainbow due to the coupling to the excited state of $^{12}$C observed in  
$^{16}$O+$^{12}$C  rainbow scattering is not seen in the calculated angular distributions.
In this sense,  $^{16}$O+$^{14}$C rainbow scattering is similar to the  $^{16}$O+$^{16}$O
 system \cite{Khoa2000,Khoa2007} rather than the  $^{16}$O+$^{12}$C system \cite{Ohkubo2014}
 in the way that $\alpha$+$^{14}$C  scattering \cite{Michel1993} is similar to 
$\alpha$+$^{16}$O scattering \cite{Michel1998}.

%fig4 Airy elephant
\begin{figure}[t]
\includegraphics[keepaspectratio,width=8.7cm] {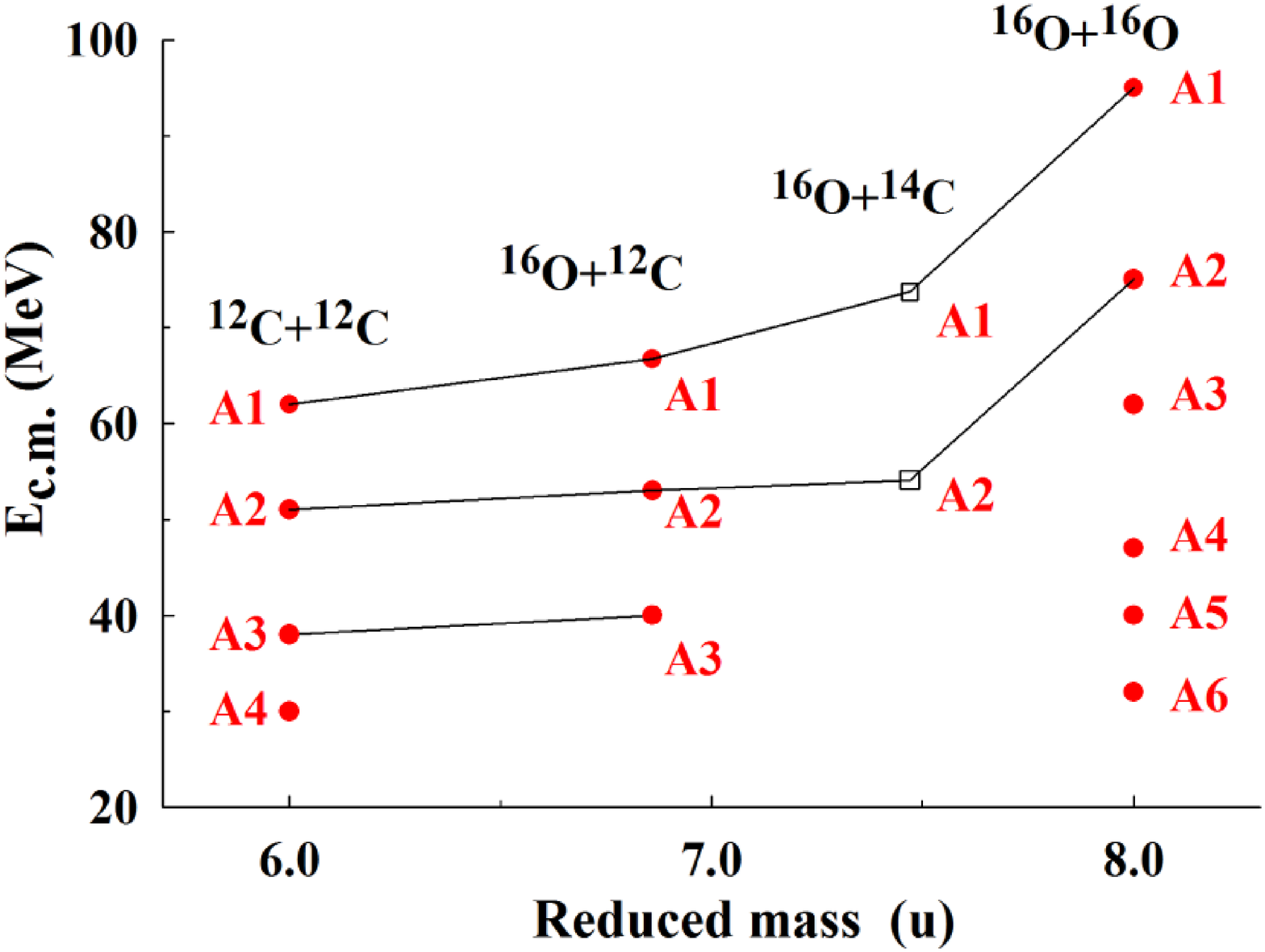}% Here is how to import EPS art
 \protect\caption{\label{fig.4} {(Color online) 
The observed  minima (filled circles) in the  90$^\circ$ excitation 
functions for the    $^{16}$O+$^{16}$O \cite{Ohkubo2002,Halbert1974,Nicoli1999,Khoa2000}, 
  $^{12}$C+$^{12}$C  \cite{McVoy1992,Michel2004,Reilly1973}, and   
 $^{16}$O+$^{12}$C \cite{Ogloblin2000} systems
are shown as a function of the reduced mass. 
The predicted Airy minima for the $^{16}$O+$^{14}$C system, $A1$ and $A2$, 
are  indicated by    open squares. The line is to guide the eye.
}
}
\end{figure}

%fig.5  volume integral energy dependence comparison 
\begin{figure}[b]
\includegraphics[keepaspectratio,width=8.7cm] {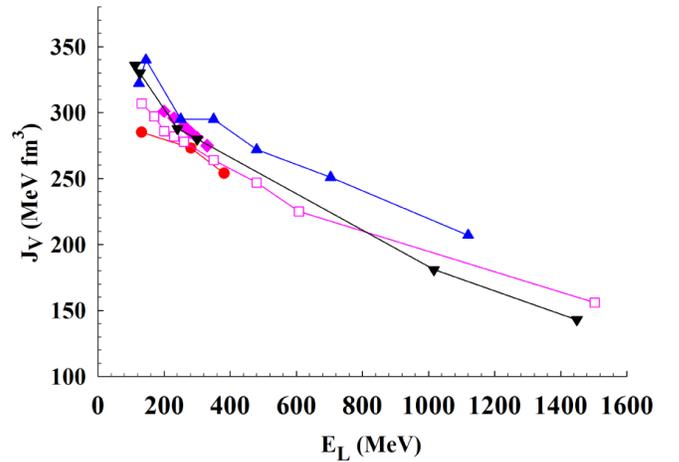}% Here is how to import EPS art
 \protect\caption{\label{fig.5} {(Color online) 
 The values of the volume integrals per nucleon pair of the real potential, $J_V$, for 
 $^{16}$O+$^{14}$C rainbow scattering (red filled circle) are shown in comparison with 
those for  
  $^{16}$O+$^{12}$C (pink open  square \cite{Ogloblin2000}, pink filled diamonds \cite{Ohkubo2014}),
  $^{16}$O+$^{16}$O (blue upper triangle) \cite{Khoa2000} and  
  $^{12}$C+$^{12}$C (black down triangle)  \cite{Khoa1994} rainbow scattering.
The line is to guide the eye.
}
}
\end{figure}

\section{AIRY MINIMA AND  AIRY ELEPHANTS IN THE  $90^\circ$  EXCITATION FUNCTION} 
\par 
Airy elephants  in the $90^\circ$  excitation function in heavy-ion 
 rainbow scattering  has been a continuing interest \cite{Khoa2007,Demyanova2010} since
 their famous discovery  in the $^{12}$C+$^{12}$C excitation function
 \cite{McVoy1992}. The existence of the Airy elephants and their numbers  can be determined 
calculating     the Airy
 minima that cross the $90^\circ$ excitation function. 
 To determine the  Airy minima theoretically, the global interaction 
potential that describes rainbow scattering over a wide range of incident energies 
has to be determined uniquely.   The  energy of  the 
$A1$ minimum  in the  $90^\circ$ excitation function  can
 be determined  using the global  potential.   In the $^{16}$O+$^{16}$O 
system, which has been  most thoroughly investigated,    its unique global potential  
has made it possible to understand   Airy Structure \cite{Michel2001},   molecular 
 resonances and cluster structure with the $^{16}$O+$^{16}$O configuration 
at low energy in a unified way \cite{Ohkubo2002}. 
As seen in Fig.~4, the $A1$ Airy minimum in the $90^\circ$  excitation function  appears at around 
$E_{c.m.}$=95 MeV and  other higher order Airy minima 
 $A2$, $A3$, $A4$, $A5$ and  $A6$ appear at around $E_{c.m.}$= 75,
 62, 47, 40 and 32 MeV, respectively \cite{Ohkubo2002,Michel2001,Khoa2000,Nicoli1999}.
The highest order of the Airy minimum is $A6$ 
 for the $^{16}$O+$^{16}$O system.
 At the lower energies below $E_{c.m.}$=32 MeV, the gross structure of the Airy structure 
evolves  into  the gross structure of the molecular resonances with the  $^{16}$O+$^{16}$O
 structure as was  shown in Ref.\cite{Ohkubo2002}. 
In Fig.~4, the observed  $A1$ Airy minimum and the highest 
order Airy minimum for the  $^{12}$C+$^{12}$C system determined from 
Refs.\cite{McVoy1992,Michel2004,Reilly1973} and  those for the 
 $^{16}$O+$^{12}$C system from Ref.\cite{Ogloblin2000} are also displayed.
For the $^{16}$O+$^{12}$C system
the  energy of  the $A1$ Airy minimum was interpolated from the experimental result at $E_L$=170 MeV
 ($E_{c.m.}$=72.9 MeV)  and  132 MeV  ($E_{c.m.}$=56.6 MeV) in Ref.\cite{Ogloblin2000}.
In Fig.~4 there exist four  Airy  minima  for $^{12}$C+$^{12}$C system and
 three Airy minima for the  $^{16}$O+$^{12}$C system.
 For the $^{16}$O+$^{14}$C system, although there is no experimental data, the energy
 evolution  of the Airy minimum in Fig.~2  predicts that  the $A1$ minimum  appears 
at  $90^\circ$ at  $E_L$=158 MeV and the  $A2$ minimum  appears at   $E_L$=116 MeV. 
 We see in Fig.~4  that the $^{16}$O+$^{14}$C system 
is situated  between the  two  identical systems, $^{16}$O+$^{16}$O and  $^{12}$C+$^{12}$C.
 The energy between the $A1$ and $A2$ minima of about  20 MeV is similar to that of the 
 $^{16}$O+$^{16}$O  system  rather than the $^{16}$O+$^{12}$C system. From this similarity,
  Airy minima with orders higher  than 
  $A3$ are  likely to   exist below  $E_L$=115 MeV before the  transition into the molecular 
resonances with the $^{16}$O+$^{14}$C structure that  have been observed in the
 $E_{c.m.}$=30 MeV region \cite{Abbondanno1990,Freeman1992}.

\par 
In Fig.~5 the  energy evolution of the volume integral of the real
 potential  for the $^{16}$O+$^{14}$C system is compared with the systematic data
  for the $^{16}$O+$^{16}$O,  $^{12}$C+$^{12}$C  and  $^{16}$O+$^{12}$C systems.
The  volume integrals for the  $^{16}$O+$^{14}$C system are  consistent  with the 
behavior  of the other systems  in the energy region where  experimental  data are
 available.  It seems that the number of the Airy minima for  identical systems is
larger than that for  the asymmetric systems. 
It is highly desired to observe the Airy minima $A2$ and
 $A3$ for the  $^{16}$O+$^{14}$C system experimentally.
 The lowest energy (highest order) Airy structure, Airy elephant,  will evolve into the molecular resonances with the  $^{16}$O+$^{14}$C 
cluster structure in the lower energy region similar to the $^{16}$O+$^{16}$O system \cite{Ohkubo2002}.  
 The molecular resonance with the $^{16}$O+$^{14}$C  structure  has been  studied  
   theoretically \cite{Heenen1979} and experimentally \cite{Abbondanno1990,Freeman1992}.
The existence of the molecular resonances  with  18$^+$, $20^+$ and  $22^+$
(or  20$^+$, $22^+$ and  $24^+$) at $E_{c.m.}$=23.4, 27.4 and  31.05 MeV, respectively,
 have been reported by Freeman {\it et al.}\cite{Freeman1992}.  
Abbondanno {\it et al.} reported  the existence of the molecular resonances  with 
$L$ = 11, 13, 17 and (15) at $E_{c.m.}$ = 18.2, 19.1, 22.9 and 23.8 MeV, respectively 
\cite{Abbondanno1990}.  Therefore it is expected  that the gross structure, Airy elephant,
 corresponding to the fourth or fifth Airy minimum in the    90$^\circ$ excitation 
function     evolves  into  molecular resonance at around $E_{c.m.}$=30  MeV. The experimental
 study  of $^{16}$O+$^{14}$C   elastic scattering in the energy region below
 $E_L$=132 MeV and above 65 MeV is highly desired to connect the Airy structures 
(Airy elephants) and the 
 molecular resonances  with the $^{16}$O+$^{14}$C  configuration.

\section{SUMMARY}
 To summarize, 
 we studied the Airy structure in $^{16}$O+$^{14}$C rainbow scattering 
with an extended double folding 
(EDF) model that describes all the diagonal and off-diagonal coupling potentials 
derived from  the microscopic  wave functions for $^{16}$O  
  using  a density-dependent   nucleon-nucleon force.
The experimental angular distributions at $E_L$=132, 281 and 382.2 MeV  were analyzed  and 
 well reproduced by the  theoretical calculations.  The 
 Airy minimum at  $\theta$=76$^\circ$ in the angular distribution at  $E_L$=132 MeV 
was found to be a second order Airy minimum $A2$. The  number of the Airy minima
 in the 90$^\circ$ excitation function was investigated in comparison with  
the typical identical $^{16}$O+$^{16}$O and $^{12}$C+$^{12}$C systems and  at least  
two Airy minima, Airy elephants, are predicted to exist above $E_L$=110 MeV ($E_{c.m.}$=51 MeV).
 The evolution 
 of the  Airy minima in the 90$^\circ$ excitation function related to the Airy elephants	
 into  molecular resonances with the 
$^{16}$O+$^{14}$C cluster structure  in the low  energy region around $E_{c.m.}$=30 MeV
is discussed.

One of the authors (SO) thanks the Yukawa Institute for Theoretical Physics,
Kyoto University  for
 the hospitality extended  during  stays in  2015.

\end{document}